\documentclass[aip,cha]{revtex4-1}

\usepackage{graphicx}
\usepackage{amsmath,amssymb}
\usepackage{amscd,float,array,bm}
\usepackage{xcolor}
\usepackage{verbatim}

\newcommand{\Ree}{{\rm Re}}
\newcommand{\Ha}{{\rm Ha}}
\newcommand{\Rm}{{\rm Rm}}
\newcommand{\Pm}{{\rm Pm}}

\newcommand{\er}{{\bf\hat e_r}}
\newcommand{\et}{{\bf\hat e_\theta}}

\newcommand{\ez}{{\bf\hat e_z}}
\newcommand{\ve}{{\mathbf{v}}}
\newcommand{\bv}{{\bf B}}
\newcommand{\bvo}{{\bf B_0}}

\newcommand{\rv}{\ensuremath{\mathbf{r}}}

\newcommand{\lp}{\ensuremath{\left(}}
\newcommand{\rp}{\ensuremath{\right)}}

\newcommand*{\citen}[1]{%
  \begingroup
    \romannumeral-`\x 
    \setcitestyle{numbers}%
    \cite{#1}%
  \endgroup   
}

\begin{document}


\title{Chaotic wave dynamics in weakly magnetised spherical Couette flows} 



\author{Ferran Garcia}
\email[]{f.garcia-gonzalez@hzdr.de}
\affiliation{Helmholtz-Zentrum Dresden-Rossendorf, Bautzner
  Landstra\ss e 400, D-01328 Dresden, Germany.}
\affiliation{Anton Pannekoek Institute for Astronomy, University of
  Amsterdam, Postbus 94249, 1090 GE Amsterdam, The Netherlands.}
\author{Martin Seilmayer}
\author{Andr\'e  Giesecke}
\author{Frank Stefani}
\affiliation{Helmholtz-Zentrum Dresden-Rossendorf, Bautzner
  Landstra\ss e 400, D-01328 Dresden, Germany.}



\date{\today}

\begin{abstract}
Direct numerical simulations of a liquid metal filling the gap between
two concentric spheres are presented. The flow is governed by the
interplay between the rotation of the inner sphere (measured by the
Reynolds number $\Ree$) and a weak externally applied axial magnetic
field (measured by the Hartmann number $\Ha$). By varying the latter a
rich variety of flow features, both in terms of spatial symmetry and
temporal dependence, is obtained. Flows with two or three independent
frequencies describing their time evolution are found as a result of
Hopf bifurcations. They are stable on a sufficiently large interval of
Hartmann numbers where regions of multistability of two, three and
even four types of these different flows are detected. The temporal
character of the solutions is analysed by means of an accurate
frequency analysis and Poincar\'e sections. An unstable branch of
flows undergoing a period doubling cascade and frequency locking of
three-frequency solutions is described as well.
\end{abstract}


\maketitle 

\section{Lead Paragraph}

\textbf{One of the paradigms of magnetohydrodynamic flows in spherical
  geometry is the magnetised spherical Couette flow. An electrically
  conducting liquid is confined between two differentially rotating
  spheres and is subjected to a magnetic field.  Despite its
  simplicity, this model gives rise to a rich variety of
  instabilities, and it is also important from an astrophysical point
  of view. The present study advances the knowledge of the dynamics of
  this problem by describing it in terms of dynamical systems theory,
  a rigorous mathematical way to understand time dependent behaviour
  of natural systems.}

\section{Introduction}
\label{sec:int}

The problem under consideration is the magnetised spherical Couette
(MSC) flow which describes the motion of an electrically conducting
liquid confined between two differentially rotating spheres and
subjected to a magnetic field parallel to the axis of rotation.  It is
formulated in terms of the three-dimensional incompressible
Navier-Stokes equations with enforced differential rotation between
the rigid boundaries. This allows the appearance of thin shear layers
(Stewartson layer, Ref.~\citen{Ste66}) parallel to the rotation axis, along
the tangent cylinder, and thin Ekman-Hartmann boundary
layers~\cite{DoSo07}. The latter emerge in case of no-slip boundary
conditions, in order to model the set up of laboratory
experiments. The numerical treatment is thus extremely challenging
because of the high spatial resolution required to resolve these thin
layers.

The MSC system is interesting as a simple model to study how magnetic
fields interact with conducting liquids in rotating spherical
containers. Such interactions are important for the understanding of
planetary or stelas dynamos, as referenced by
Refs.~\citen{Jon11,Rud89}, and related experiments conducted by
Ref.~\citen{GLPGS02}. See the review article (Ref.~\citen{BrSu05}) and
books (Refs.~\citen{DoSo07,MoDo19}) for detailed references and
introduction to the field. The MSC system is also of interest to model
the magnetorotational instability (MRI)~\cite{BaHa91} which is
believed to explain the transport of angular momentum in accretion
disks around black holes and stars, and also in protoplanetary
disks~\cite{JiBa13}. The transport of angular momentum in such
environments seems to be the cause which allows the matter to fall
into the center.

Because of its relevance, MRI has been studied experimentally, with
GaInSn between two rotating cylinders at Helmholtz-Zentrum
Dresden-Rossendorf (HZDR)~\cite{SGGRSSH06,SGGHPRS09,SGGGSGRSH14}, and
in Maryland~\cite{SMTHDHAL04} with liquid sodium between
differentially rotating spheres. In this latter experiment the
observed instabilities were considered as MRI, but with
nonaxisymmetric (azimuthally dependent) symmetry. However, subsequent
numerical simulations~\cite{Hol09,GJG11}, and
experiments~\cite{KKSS17}, have not interpreted the instabilities as
MRI but as typical instabilities associated to the MSC system, namely
the radial jet, return flow and shear layer
instabilities~\cite{TEO11,GaSt18}. These different interpretations
provide a motivation for advancing the knowledge of the MSC system.

In absence of magnetic field, and with the outer sphere at rest, the
basic axisymmetric spherical Couette (SC) flow is stable~\cite{Sch86}
for sufficiently small rotation of the inner sphere (measured by the
Reynolds number $\Ree$). At a certain critical $\Ree_c$
nonaxisymmetric instabilities give rise to regular flows whose pattern
and topology depend strongly on the gap width
$\chi=r_{\mathrm{i}}/r_{\mathrm{o}}$ (with $r_{\mathrm{i}}$ and
$r_{\mathrm{o}}$ being the inner and outer radii) of the problem. When
the shell is thin, Taylor-G\"ortler vortices
develop~\cite{Zik96,YaLu02,Yua12} contrasting with the spiral
waves~\cite{WER99,HJE06,AYS18} preferred at wider gaps. In the latter
case, triadic resonances were found in a recent study (see
Ref.~\citen{BTHW18}) before reaching the turbulent regime. Generally,
chaotic and even turbulent flows are obtained from successive
bifurcations of the initial base state if $\Ree$ is sufficiently
increased. This is typical in large scale dissipative dynamical
systems~\cite{Eck81}. Although the SC system has a simple formulation,
it reveals immense complexity as it has been shown in the recent study
(Ref.~\citen{Wic14}) which investigates different flow regimes
appearing either with positive or negative differential rotation.

In the presence of a magnetic field the axisymmetric base spherical
Couette flow and its corresponding nonaxisymmetric instabilities have
been determined in the $(\chi,\Ree,\Ha)$ parameter
space~\cite{Hol09,TEO11}. Only a few fully nonlinear studies (e.g.,
Refs.~\citen{HoSk01,WeHo10,GJG11,FSNS13,Kap14}) have investigated the
flow properties, by means of direct numerical simulations (DNS), and
addressed their dependence on the physical parameters. The main
differences among the latter studies rely on the topology of the
applied magnetic field (axial, dipole, quadrupole) or on the type of
boundary conditions considered. A comprehensive numerical study,
Ref.~\citen{KNS18}, related to the liquid sodium Derviche Torneur
Sodium (DTS) experiment in Grenoble~\cite{Bri_et_al11}, has revealed
several dynamical regimes where coherent structures coexist with
turbulent flows. Very recently on Ref.~\citen{MLJ19}, with an inner
sphere rotating slightly faster than the outer and a dipolar field due
to magnetised inner sphere, axisymmetric and nonaxisymmetric solutions
were obtained. The latter have been interpreted as MRI because they
are produced within the Stewartson layer and arise as a rotating wave
when the helical component of the magnetic field is strong.

The HEDGEHOG (\underline{H}ydromagnetic \underline{E}xperiment with
\underline{D}ifferentially \underline{G}yrating sph\underline{E}res
\underline{HO}lding \underline{G}aInSn) laboratory
experiment~\cite{KKSS17} at Helmholtz-Zentrum Dresden-Rossendorf is
designed to study the different flow (hydrodynamic and magnetic)
instabilities occurring for a homogeneous axial applied magnetic
field. This configuration has been previously addressed in the
numerical studies of Refs.~\citen{HoSk01,Hol09,TEO11,Kap14}. Beyond a
critical value $\Ree_{\text{c}}$ the basic flow becomes unstable to
non-axisymmetric perturbations whose topology depends strongly on the
applied magnetic field strength (measured by the Hartman number
$\Ha$). At low $\Ha$ the instability appears in the form of an
equatorially antisymmetric radial jet at the equatorial plane, whereas
at large $\Ha$ the instability is equatorially symmetric and connected
to a shear layer at the tangent cylinder~\cite{HoSk01,Hol09}.  For
moderate values of $\Ha$, between the radial jet and the shear layer
instability, it takes the form of a meridional return flow
instability~\cite{Hol09}. We note that the radial jet and return flow
instabilities are separated by a $\Ha$ interval in which the basic
flow stabilises again~\cite{Hol09,TEO11}.

From the mathematical point of view the SC and the MSC are
{\bf{SO}}$(2)\times${\bf{Z}}$_2$ equivariant systems, i.\,e.,
invariant by azimuthal rotations and reflections with respect to the
equatorial plane. In this class of dynamical systems the type of
solutions appearing at successive bifurcations from the base state is
theoretically known~\cite{Ran82,CrKn91,EZK92,GLM00,GoSt03}.

After a primary Hopf bifurcation, several branches of periodic
rotating waves (RW) appear, which in turn give rise to quasiperiodic
modulated rotating waves (MRW) at a secondary Hopf bifurcations. In
case of the MSC problem the existence of RW has been confirmed by
experimental studies~\cite{SABCGJN08} and by
DNS~\cite{HoSk01,Hol09,GJG11} but their dependence upon parameters,
especially the Hartmann number, was not addressed in much detail. Only
very recently, a continuation method was applied in
Ref.~\citen{GaSt18} to determine extensive RW bifurcation diagrams as
a function of $\Ha$. The stability analysis of RW close to secondary
bifurcation points have provided initial conditions to obtain and
classify MRW~\cite{GSGS19}. The analysis of
Refs.~\citen{GaSt18,GSGS19} is performed at a moderate Reynolds number
regime extending previous numerical studies~\cite{HoSk01,Hol09,TEO11}
by describing the type of flows in terms of bifurcation theory.
  
The pioneering theoretical work of Ref.~\citen{NRT78} showed that
arbitrarily smooth perturbations can destroy a three-frequency
quasiperiodic solution, a solution whose temporal dependence is
described by three incommensurate frequencies, producing a strange
attractor questioning the existence of such a solutions in real
physical systems. Later work of Ref.~\citen{GOY83}, however, provided
several numerical examples where these three frequency solutions
indeed exist. The authors of Ref.~\citen{GOY83} argue that the smooth
perturbations which may destroy these three frequency solutions are
unlikely to occur in practise. Subsequent studies in the context of
axisymmetric Navier-Stokes flow in a cylindrical
annulus~\cite{LoMa00}, 2D convection in a rectangle~\cite{BKMN01}, a
rotating baroclinic annulus~\cite{RFRM06}, or a 2D rectangular cavity
filled with air~\cite{ODPQ15}, provided further evidence of three
frequency solutions. Moreover, three frequency solutions have been
found as well as in the case of 3D convection in a rotating sphere,
although in a very narrow parameter range~\cite{GNS16}.

The present paper continues previous work described in
Refs.~\citen{GaSt18,GSGS19}. The main novelty of the present study is
to describe new branches of quasiperiodic MRW with two and even three
frequencies and branches of chaotic flows. The flows with temporal
dependence described by 3 frequencies are obtained for first time in
the MSC problem. In addition, to the best knowledge of the authors,
three frequency solutions in a three dimensional MHD problem have been
never described up to now.  These solutions are found when the applied
axial magnetic field is weak, i.\,e., in the regime governed by the
radial jet instability. The main goal is to describe the
spatio-temporal symmetries of the waves as the Hartmann number
$\Ha\rightarrow 0$ is decreased. As it will be shown in the paper,
several regions of multistability involving three-frequency solutions
are found and their temporal scales are accurately determined by means
of sophisticated time series analysis~\cite{Las93}. This is important
for later comparisons with the results of the HEDGEHOG experiment.

The organisation of the paper is as follows: Section \S\ \ref{sec:mod}
introduces the model problem and the numerical method used to solve
the governing equations. The core of the study belongs to
\S\ \ref{sec:res} where the results are presented and the dynamics of
the system is analysed. Finally, \S\ \ref{sec:disc} summarises the
main conclusions of the study.

\section{The Magnetised Spherical Couette Model}
\label{sec:mod}

A liquid metal of constant density $\rho$, kinematic viscosity $\nu$,
electrical conductivity $\sigma$, magnetic diffusivity
$\eta=1/(\sigma\mu_0)$, $\mu_0$ being the free-space value for the
magnetic permeability, fills a spherical shell with inner and outer
radii $r_{\mathrm{i}}$ and $r_{\mathrm{o}}$.  The fluid is driven by
the rotation of the inner sphere at a constant angular velocity
$\Omega$ around the vertical axis $\ez$ and the outer sphere is at
rest.

\begin{figure*}[h!]
\begin{center}
  \includegraphics[scale=1.]{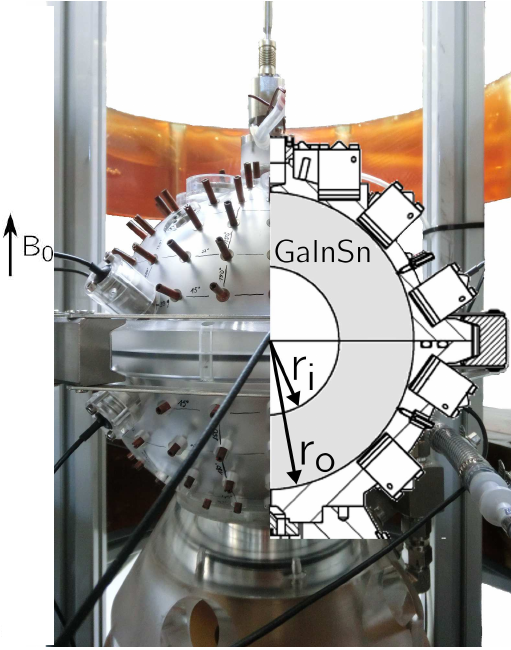}
\end{center}  
\caption{Geometry of the HEDGEHOG experiment. The scheme (left side)
  shows the inner (rotating) and outer (at rest) spheres. On the
  latter, the ultrasonic Doppler velocimeter (UDV) sensors (thick
  cylinders) and the electric potential probes (thin needles) are
  attached.}
\label{fig:hed_exp}   
\end{figure*}

In the case of the HEDGEHOG laboratory experiment, compare
to Ref.~\citen{KKSS17}, the flow is subjected to a uniform axial magnetic
field $\bvo=B_0\ez=B_0 \cos(\theta)\er-B_0 \sin(\theta)\et$, $\theta$
being the colatitude and $B_0$ the magnetic field strength
(Fig.~\ref{fig:hed_exp}).

Length, time, velocity and magnetic field are scaled using the
characteristic quantities $d=r_{\mathrm{o}}-r_{\mathrm{i}}$, $d^2/\nu$, $r_{\mathrm{i}}\Omega$ and
$B_0$, respectively, in order to obtain the dimensionless
equations. For the HEDGEHOG experiment the inductionless approximation
$\Rm\ll 1$ is valid as the parameter regime is in the limit of low
magnetic Reynolds number $\Rm=\Omega r_{\mathrm{i}} d/\eta$. This is because the
fluid has very low magnetic Prandtl number $\Pm=\nu/\eta\sim
O(10^{-6})$ (eutectic alloy GaInSn~\cite{PSEGN14}) and only moderate
Reynolds numbers $\Ree=\Omega r_{\mathrm{i}} d/\nu\sim 10^3$ are considered
giving rise to $\Rm=\Pm\Ree \sim 10^{-3}$.

Decomposing the magnetic field as $\bv=\ez+\Rm{\bf b}$ and
neglecting terms $O(\Rm)$, the Navier-Stokes and induction equations read
\begin{align}
  \partial_t\ve+\Ree\lp\ve\cdot\nabla\rp\ve &=
-\nabla p+\nabla^2\ve+\Ha^2(\nabla\times {\bf b})\times\ez, \label{eq:mom_less}   \\
 0& = \nabla\times(\ve\times\ez)+\nabla^2{\bf b}, \label{eq:ind_less}\\
\nabla\cdot\ve=0, &\quad \nabla\cdot{\bf b}=0.\label{eq:div}
\end{align}
In the framework of the inductionless approximation the MSC system is
described in terms of three non-dimensional numbers: the Reynolds
number, the Hartmann number and the aspect ratio
\begin{equation*}
  \Ree=\frac{\Omega r_{\mathrm{i}} d}{\nu}, \quad
  \Ha=\frac{B_0d}{\sqrt{\mu_0\rho\nu\eta}}=B_0d\sqrt{\frac{\sigma}{\rho\nu}},\quad
  \chi=\frac{r_{\mathrm{i}}}{r_{\mathrm{o}}}.
\end{equation*}
The boundary conditions for the velocity field are no-slip
($v_r=v_\theta=v_\varphi=0$) at $r=r_{\mathrm{o}}$ and constant
rotation ($v_r=v_\theta=0,~v_\varphi= \sin{\theta}$) at
$r=r_{\mathrm{i}}$. As in previous studies (e.g.,
Ref.~\citen{HoSk01}), and mimicking the experimental setting,
insulating exterior regions are considered for the magnetic field.

The method to numerically solve the equations is described in
Ref.~\citen{GaSt18} and references therein. The toroidal, $\Psi$, and
poloidal, $\Phi$, potentials are utilised to express the
divergence-free velocity field
\begin{equation}
  \ve=\nabla\times\lp\Psi\rv\rp+\nabla\times\nabla\times\lp\Phi\rv\rp.
\label{eq:pot}  
\end{equation}
The unknowns are expanded in spherical harmonics in the angular
coordinates ($\rv=r~\er$ is the position vector) and a collocation
method on a Gauss--Lobatto mesh of $N_r$ points is considered for the
radial direction.  More concretely, the solution vector
$u=(\Psi,\Phi)$ (Eq.~\ref{eq:pot}) is expanded in spherical harmonic
series up to degree $L_{\text{max}}$ and order
$M_{\text{max}}=L_{\text{max}}$ as
\begin{eqnarray}
  \Psi(t,r,\theta,\varphi)=\sum_{l=0}^{L_{\text{max}}}\sum_{m=-l}^{l}{\Psi_{l}^{m}(r,t)Y_l^{m}(\theta,\varphi)},\label{eq:serie_psi}\\
  \Phi(t,r,\theta,\varphi)=\sum_{l=0}^{L_{\text{max}}}\sum_{m=-l}^{l}{\Phi_{l}^{m}(r,t)Y_l^{m}(\theta,\varphi)},\label{eq:serie_phi}
\end{eqnarray}
with $\Psi_l^{-m}=\overline{\Psi_l^{m}}$,
$\Phi_l^{-m}=\overline{\Phi_l^{m}}$, $\Psi_0^0=\Phi_0^0=0$ to uniquely
determine the two scalar potentials, and
$Y_l^{m}(\theta,\varphi)=P_l^m(\cos\theta) e^{im\varphi}$, $P_l^m$
being the normalised associated Legendre functions of degree $l$ and
order $m$. The code is parallelised on the spectral and on the
physical space by using OpenMP directives. Optimised libraries
(FFTW3~\cite{FrJo05}) for the FFTs in $\varphi$, and matrix-matrix
products (dgemm GOTO~\cite{GoGe08}) for the Legendre transforms in
$\theta$, are implemented for the computation of the nonlinear
(advection) term.

High order implicit-explicit backward differentiation formulas
(IMEX--BDF)~\cite{GNGS10} are used for the time integration. In the
IMEX method we treat the nonlinear terms explicitly in order to avoid
solving a nonlinear system of equations at each time step. The Lorenz
force term in Eq.~\ref{eq:mom_less} is treated explicitly too, which
may necessitate a reduced time step in comparison with an implicit
treatment. However, this is not a serious issue when moderate $\Ha$
are considered, as is the case for the present study.  The use of
\textit{matrix-free} Krylov methods (GMRES~\cite{SaSc86} in our case)
for the linear systems facilitates the implementation of a suitable
order and time stepsize control for the time integration (see
Ref.~\citen{GNGS10} for details of the implementation).

\section{Results}
\label{sec:res}

The results presented in this section are based on direct numerical
simulations for fixed $\chi=0.5$, $\Ree=10^3$ and control parameter
$\Ha\in[0,6]$. According to table 1 of Ref.~\citen{GaSt18} this
parameter regime is well resolved for $N_r=40$ (number of radial
collocation points) and $L_{\text{max}}=84$ (spherical harmonics
truncation parameter). From time to time, we increase the resolution
to $N_r=60$ and $L_{\text{max}}=126$ to seek for discretisation errors
which are below $1\%$.

The two previous studies (Refs.~\citen{GaSt18,GSGS19}) provided the
bifurcation diagrams for rotating waves and modulated rotating waves
including the above mentioned range of parameters. Specifically,
branches of unstable/stable rotating waves with azimuthal symmetry
$m=2,3,4$ were computed~\cite{GaSt18} and branches of MRW with
azimuthal symmetries $m=1,2,3$ were classified~\cite{GSGS19} according
to the established theory~\cite{Ran82,GLM00}. Here, we start to
compute the new flows using DNS with initial conditions built from RW
or MRW previously obtained~\cite{GaSt18,GSGS19}. By adding a small
random perturbation to all the spherical harmonic amplitudes of these
initial conditions, the model equations are time-stepped until an
attractor is reached, that is, the flow is saturated and in a
statistically steady state. To study the dependence of the flows on
$\Ha$ we analyse time series of local (the radial velocity measured at
a point inside the shell) or global physical properties such as the
volume-averaged kinetic energy $K$. The latter is defined as
\begin{equation}
K=\frac{1}{2{\mathcal V}}\int_{\mathcal V} \ve
\cdot \ve \;dv.
\label{eq:ener_dens}
\end{equation}
Concretely, we consider this volume integral for the non-axisymmetric
component of the velocity field to define the non-axisymmetric
$K_{na}$ kinetic energy. It is based on the $m\neq 0$ modes of the
spherical harmonic expansion of the potentials $\Psi$ and $\Phi$. This
choice is motivated by the non-axisymmetric character of the radial
jet instability, which is the focus of the present study. In addition,
the previous volume integral can be computed for either the toroidal,
$\bm\nabla\times\lp\Psi\rv\rp$, or the poloidal,
$\bm\nabla\times\bm\nabla\times\lp\Phi\rv\rp$, component of the
velocity field, giving rise to either the toroidal, $K^{\text{T}}$, or
the poloidal, $K^{\text{P}}$, kinetic energies.

For each wave number $m$ in the spherical harmonic expansion
(Eqs.~\ref{eq:serie_psi}-\ref{eq:serie_phi}), the instantaneous
kinetic energy $K_m$ is defined by only considering the spherical
harmonic amplitudes $\psi^m_l$ and $\phi^m_l$ with order $m$ and
degree $|m|\le l \le L_{\text{max}}$. For each DNS, we search for
$m_{\text{max}}$ with $\overline{K}_{m_{\text{max}}}\ge
\overline{K}_m$, $1\le m\le L_{\text{max}}$, $\overline{·}$ being the
time average. In our implementation an $m$-fold azimuthal symmetry for
fixed $m=m_d$ can be imposed to obtain solutions with prescribed
symmetry. This means that only the spherical harmonic amplitudes with
azimuthal wave numbers being multiples of $m_d$ are nonzero in
Eqs.~\ref{eq:serie_psi}-\ref{eq:serie_phi}. We recall that a solution
with $m$-fold azimuthal symmetry is unaffected by azimuthal rotations
multiple of $2\pi/m$. Notice that if the azimuthal symmetry is $m=1$
all the spherical harmonics amplitudes are considered. Flows with
azimuthal symmetry $m=1$ are considered stable if their time
integration, after filtering initial transients, remains on the same
attractor for sufficiently large time. In the case of flows obtained
by constraining the azimuthal symmetry ($m=2$ or $m=3$) their
stability is assessed by adding an $m=1$ random perturbation. They are
considered unstable if the $m=1$ perturbation grows in time. By means
of time spectrum analysis and Poincar\'e sections we infer the
quasiperiodic or chaotic character of the DNS.

\subsection{Type of flows with  $\Ha<6$}
\label{sec:typ_flow}

Because the basic flow is axisymmetric ($m=0$ azimuthal symmetry) the
volume-averaged nonaxisymmetric ($m\ne0$) component of the kinetic
energy, $K_{\text{na}}$, is used as a proxy for the bifurcation
diagrams of the time dependent flows presented in this
section. Concretely, the time average of the time series of
$K_{\text{na}}$ for each computed solution is considered. In addition,
we display the maximum and minimum values of the time series of
$K_{\text{na}}$ to estimate the amplitude of fluctuations. For this
purpose, sufficiently large time series (more than 10 dimensionless
time units, i.\,e. viscous time-scales) have to be obtained. This is
especially true close to bifurcation points where long initial
transients are expected.

\begin{figure*}[h!]
\begin{center}
  \includegraphics[scale=2.]{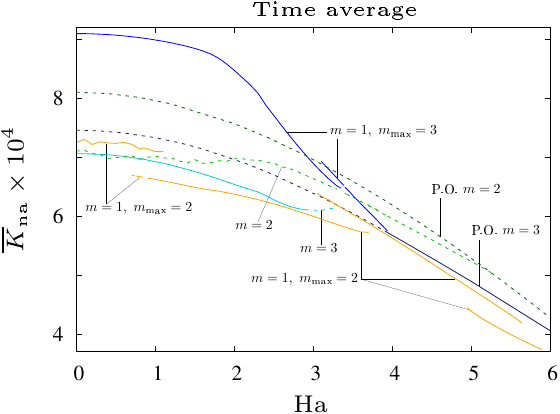}
\end{center}  
\caption{Bifurcation diagrams of the nonaxisymmetric ($m\ne0$) time
  and volume averaged kinetic energy $\overline{K}_{\text{na}}$ versus
  the Hartmann number $\Ha$ for weakly magnetised flows corresponding
  to the equatorially asymmetric radial jet instability (color
  online). The azimuthal symmetry $m$ and most energetic azimuthal
  wave number $m_{\text{max}}$ are displayed on each branch. Four
  general classes of complex flows are shown: flows with $m=1$ and
  $m_{\text{max}}=3$ (blue), flows with $m=1$ and $m_{\text{max}}=2$
  (orange), flows with $m=3$ (turquoise) and flows with $m=2$
  (green). Branches of periodic orbits with $m=2,3$ are displayed as
  well. Solid/dashed lines mean stable/unstable flows.}
\label{fig:bif_diagr}   
\end{figure*}

Figure~\ref{fig:bif_diagr} displays the time average
$\overline{K}_{\text{na}}$ versus the Hartmann number $\Ha$ for weakly
magnetised flows corresponding to the radial jet instability. Several
branches are shown and two integers --the azimuthal symmetry $m$ and
most energetic azimuthal wave number $m_{\text{max}}$-- are used to
label them.  As usual, stable/unstable motions are marked with
different line type (solid/dashed). The branches shown in the figure
can be classified as four classes of flows: flows with $m=1$ and
$m_{\text{max}}=3$, flows with $m=1$ and $m_{\text{max}}=2$, flows
with $m=3$ and flows with $m=2$.

For each of the two latter cases ($m=2,3$) a branch of rotating waves
(periodic orbits labelled with P.O.), as previously computed in
Ref.~\citen{GaSt18}, is shown. Hopf bifurcations along these
P.O. branches give rise to modulated rotating waves MRW, with $m=1$
and $m_{\text{max}}=3$, $m=3$ or $m=2$, whose spatio-temporal symmetry
was analysed in Ref.~\citen{GSGS19}. The branches of complex flows
corresponding to $m=1$ and $m_{\text{max}}=2$ are however not directly
connected with those of the P.O. For this class of flows ($m=1$ and
$m_{\text{max}}=2$) some MRW where studied in Ref.~\citen{GSGS19} as
well. The MRW of Ref.~\citen{GSGS19} have been described for
$\Ha>2.5$, while in Fig.~\ref{fig:bif_diagr} their branches are
extended down to $\Ha=0$. In addition, Fig.~\ref{fig:bif_diagr}
contains other branches of waves and chaotic flows not previously
found in Ref.~\citen{GSGS19}.

In order to simplify the exposition of the description of the flows
displayed in Fig.~\ref{fig:bif_diagr} the class of flows with $m=1$
and $m_{\text{max}}=3$, and the class of flows with $m=3$ can be
joined into the class $m_{\text{max}}=3$. Then, three different
classes (i) $m_{\text{max}}=3$ (ii) $m=1$ and $m_{\text{max}}=2$, and
(iii) $m=2$, are obtained. The three following sections describe these
classes of flows, and the bifurcations among them.

\subsection{Flows with $m_{\textnormal{max}}=3$}
\label{sec:m3}

\begin{figure*}[h!]
\begin{center}
  \includegraphics[scale=1.35]{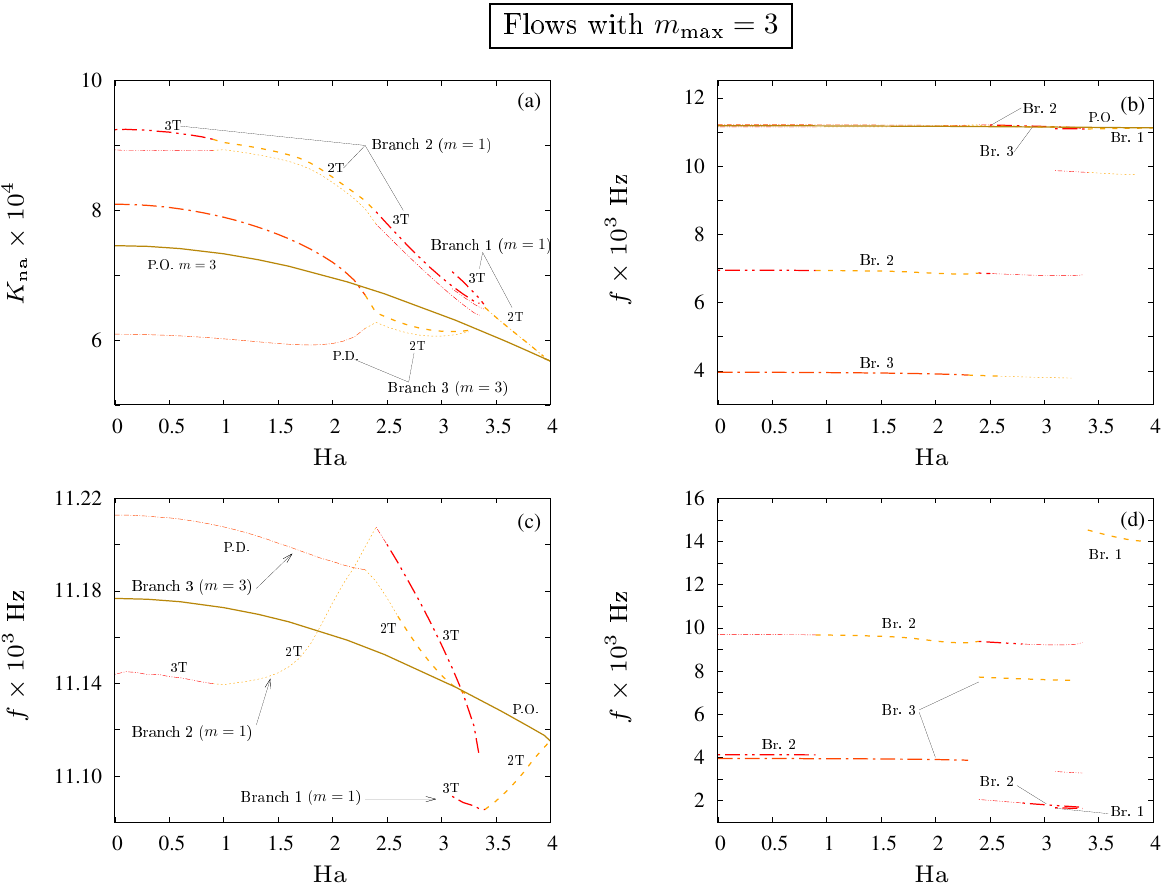}
\end{center}
\caption{Flows with $m_{\text{max}}=3$ (either with $m=1$ or
  $m=3$). (a) Maximum (thick line) and minimum (thin line) of
  $K_{\text{na}}$. (b) 1st (thick line) and 2nd (thin line) dominant
  frequencies (the two having largest amplitude) of the flow. (c)
  Detail of (b) showing the frequencies in the range of $\approx
  0.011$ Hz. (d) 1st (thick line) and 2nd (thin line) dominant
  frequencies of the volume-averaged kinetic energy.  The type of
  solutions --periodic orbit (P.O., solid line), invariant tori (2T,
  dashed line), invariant tori with period doubling (P.D.,
  dashed-dotted line), and invariant tori with three frequencies (3T,
  dashed-double-dotted line)-- is marked on each branch.}
\label{fig:bif_diagr_m3}    
\end{figure*}

Figure~\ref{fig:bif_diagr_m3}(a) details the type of temporal
dependence detected on each of the branches shown in
Fig.~\ref{fig:bif_diagr} corresponding to the class of flows with
$m_{\text{max}}=3$. The maximum (thick line) and minimum (thin line)
values of the volume-averaged nonaxisymmetric ($m\ne0$) kinetic
energy, for three different branches with $m=1$-fold azimuthal
symmetry (Branch 1 and Branch 2) and with $m=3$-fold azimuthal
symmetry (Branch 3), are plotted versus $\Ha$. Solutions of Branch 1,
and those of Branch 3 for $\Ha\in[2.5,3.25]$, were studied previously
in Ref.~\citen{GSGS19}. We recall that both branches arise, due to a
Hopf bifurcation on the branch of RW (periodic orbits, P.O., solid
line), giving rise to MRW which are invariant tori with two
frequencies (2T, dashed line). As noticed in Ref.~\citen{GSGS19} the
volume-averaged properties of MRW of Branch 1 have very small
oscillations and thus their maximum and minimum values are almost the
same (see Fig.~\ref{fig:bif_diagr_m3}(a)). This does not occur for MRW
of Branch 3, as it is usually expected for these type of
waves~\cite{GNS16}.

By decreasing $\Ha$ from $\Ha=2.5$ and taking initial conditions on
the MRW (2T) of Branch 3 a period doubling bifurcation (labelled P.D.,
dashed-dotted line) is detected and the branch extends down to
$\Ha=0$. Similarly, decreasing $\Ha$ but from MRW of Branch 1, a
branch of invariant tori with three frequencies (3T,
dashed-double-dotted line) is obtained at around $\Ha\approx 3.4$ as a
result of a Hopf bifurcation. By taking a 3T initial condition of
Branch 1 with $\Ha=3.1$ and decreasing $\Ha$ down $\Ha=3$ a new
branch, called Branch 2, is found. The type of solutions is 3T and
hysteresis behaviour is observed as this new branch can be continued
up to $\Ha\approx 3.35$. A subcritical pitchfork bifurcation, together
with a re-stabilisation and possible folding, may be the origin of
this new branch as we will argue later on the text. By extending
Branch 2 from $\Ha=3$ down to $\Ha=0$ a 2T solution, and a 3T
solution, branches are found. In other words, we have found a branch
of 2T solutions connecting two branches of 3T solutions.

Figure~\ref{fig:bif_diagr_m3}(b,c,d) display the result of the
frequency analysis in Hz. As we solve dimensionless equations, the
frequencies should be translated in Hz. We assume the kinematic
viscosity of the eutectic alloy Ga$^{67}$In$^{20.5}$Sn$^{12.5}$ may be
approximated by $\nu=3.4\times 10^{-3}$\nolinebreak cm$^2$s$^{-1}$
(see Ref.~\citen{PSEGN14}). Together with the geometric setup of the
Hedgehog experiment, with $r_{\mathrm{o}}=9$ cm and
$r_{\mathrm{i}}=4.5$ cm, the dimensional value for the frequencies
computed in this study is $f=f^*\nu/d^2=1.7\times 10^{-3}f^*$ Hz,
$f^*$ being a dimensionless
frequency. Figure~\ref{fig:bif_diagr_m3}(b) (and its detail (c)) shows
the 1st (thick line) and 2nd (thin line) dominant
Laskar's~\cite{Las93} frequencies (the two frequencies having largest
amplitude). They are computed from the time series of the radial
velocity picked up at the point $(r_{\mathrm{i}}+0.5d,\pi/8,0)$ for
solutions of the same branches of Fig.~\ref{fig:bif_diagr_m3}(a).

Figure~\ref{fig:bif_diagr_m3}(b) evidences an interchange of the two
dominant frequencies on all the main branches (Branch 1, 2, and 3) as
$\Ha$ is decreased. We have to note however that the amplitudes of
both frequencies remain similar. Because the 1st and 2nd dominant
frequencies are not related by an integer constant, they are
independent and thus describe the temporal scales (in Hz) expected for
flows at this range of parameters. As seen in
Fig~\ref{fig:bif_diagr_m3}(b) all the branches have one frequency in
the range $[11.08,11.22]$ which in turn is quite similar to the unique
frequency along the branch of P.O., which corresponds to the rotating
frequency of the waves~\cite{Ran82}. A detail of this is shown in
Fig~\ref{fig:bif_diagr_m3}(c) where the transitions among the
different type of flows can be clearly identified.

Figure~\ref{fig:bif_diagr_m3}(d) is as Fig.~\ref{fig:bif_diagr_m3}(b),
but for the time series of the volume-averaged kinetic energy. Because
it is a volume (and thus azimuthally) averaged quantity, the drifting
frequency of the flow shown in Figs.~\ref{fig:bif_diagr_m3}(b,c) does
not appear and the frequencies correspond exclusively to those driving
the modulation. This means in turn that 2T solutions exhibit only one
frequency and 3T solutions exhibit 2 frequencies. From
Fig.~\ref{fig:bif_diagr_m3}(d) it becomes clear that on Branch 3 a
period doubling bifurcation occurs around $\Ha=2.3$. Moreover, close
to $\Ha=3.35$ the frequencies of the 3T solutions at Branch 1 and at
Branch 2 are very similar as it is the case for the drift frequency
shown in Fig.~\ref{fig:bif_diagr_m3}(c). This provides some evidence
that both branches are related and, as commented earlier, the relation
could be a pitchfork bifurcation. The study of the main frequencies,
and thus the temporal scales of the flow, is relevant for future
comparisons with the HEDGEHOG experiment~\cite{KKSS17}.

\begin{figure*}[h!]
\begin{center}
  \includegraphics[scale=1.15]{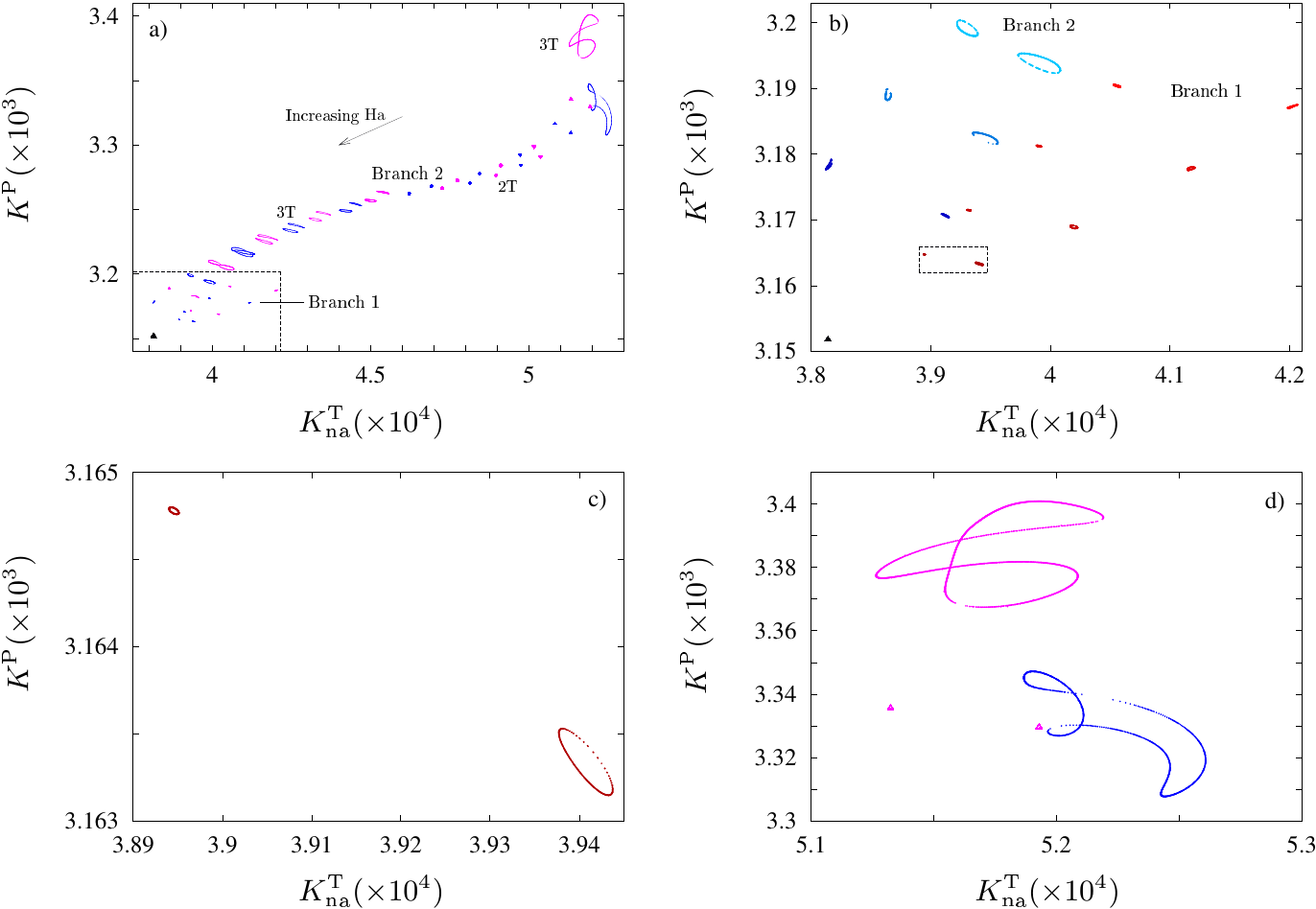}
\end{center}
\caption{Flows with $m=1,~m_{\text{max}}=3$: Poincar\'e sections
  defined by the constraint $K(t)=\overline{K}$,  $K$ being the
  volume-averaged kinetic energy and $\overline{K}$ its time
  average. The volume-averaged poloidal kinetic energy
  $K^{\text{P}}$ is displayed versus the volume-averaged
  toroidal nonaxisymmetric energy $K^{\text{T}}_{\text{na}}$. (a)
  Sequence of invariant tori with two (2T, points) and three (3T,
  closed curves) frequencies for Branches 1 and 2. The Hartmann
  numbers are $\Ha=3.1,3.2,3.3,3.37,3.5$ for Branch 1 and
  $\Ha=0,0.7,1.1,1.5,1.8,1.9,2,2.1,2.2,2.3,2.4,2.5,2.6,2.7,2.8,2.9,3.,3.1,3.2,3.3$
  for branch 2. The solution on the left bottom corner corresponds to
  a 2T solution at $\Ha=3.5$.  (b) Detail of (a) showing the
  transition between 2T solution (point on the left bottom corner) and
  the two branches of 3T solutions. (c) Detail of (b) displaying a 3T
  solution on branch 1 very close to the bifurcation. (d) Detail of
  the Poincar\'e sections of complex 3T attractors found at $\Ha=0$
  and $\Ha=0.7$ and those of a 2T solution at $\Ha=1.1$.}
\label{fig:poinc_m3}    
\end{figure*}

The analysis of the two main frequencies of the flow does not provide
a complete description of the time dependence as the time spectrum
usually consists of many of them. We now confirm that for a 2T (3T)
solution all frequencies present in the spectrum have to be an integer
linear combination of two (three) fundamental
frequencies~\cite{Las93,GMS10}. The Poincar\'e section helps to
rigorously validate the existence of 2T solution from a time series of
a dynamical system. The Poincar\'e section constructed from a periodic
time series is a point whereas it corresponds to a closed curve if the
time series involves two fundamental frequencies. In our case,
considering the time series of volume-averaged quantities (and thus
eliminating one frequency of the system) will help to identify 3T
solutions as its Poincar\'e section will correspond to a closed curve.

Figure~\ref{fig:poinc_m3} displays the (two) Poincar\'e sections
defined by the constraint $K(t)=\overline{K}$, $K$ being the
volume-averaged kinetic energy and $\overline{K}$ its time
average. The volume-averaged poloidal kinetic energy $K^{\text{P}}$ is
plotted versus the volume-averaged toroidal nonaxisymmetric energy
$K^{\text{T}}_{\text{na}}$. The figure illustrates the bifurcation
scenario on Branch 1 and 2. For each solution of
Fig.~\ref{fig:poinc_m3}(a) the two Poincar\'e sections are very close
with their Hartmann numbers increasing from the top right corner
following the arrow. For Branch 1 $\Ha\in[3.1,3.5]$, while for Branch
2 $\Ha\in[0,3.3]$. The triangle on the left bottom corner corresponds
to a 2T solution (a point) at $\Ha=3.5$ and by decreasing $\Ha$ the
two branches of 3T solutions (Branch 1 and 2) can be identified as
closed curves (see details of Fig.~\ref{fig:poinc_m3}(b,c)). As
commented before, on Branch 2 there is a branch of 2T solutions
(points on the right part of Fig.~\ref{fig:poinc_m3}(a)) connecting
the two branches of 3T solutions. The Poincar\'e sections appearing at
the right top corner (see detail Fig.~\ref{fig:poinc_m3}(d))
correspond to complex 3T attractors found at $\Ha=0$ and $\Ha=0.7$.

The solutions of Branch 1 and 2 are stable since they are obtained by
means of DNS without symmetry assumptions. Solutions on Branch 3,
obtained when imposing an $m=3$ symmetry, are unstable for
$\Ha\in[2.9,3.25]$ and they are stable for $\Ha<2.9$. The stability of
these solutions is assessed by analysing the growth or decay of a
random perturbation with azimuthal symmetry $m=1$. The interval of
Hartmann numbers for the stability of each type of temporal dependence
(P.O., 2T, and 3T) on each branch shown in
Fig.~\ref{fig:bif_diagr_m3}(a) is provided in Table~\ref{table:m3}.

\begin{table}[ht]
\caption{Range of Hartmann numbers for the stability of the different
  types of temporal dependence along the different branches (Branch
  1, 2 and 3) for flows with $m_{\text{max}}=3$. The interval of
  definition of the nonaxisymmetric radial jet instability is
  $\Ha\in[0,\Ha_c)$ with $\Ha_c=12.2$ being the critical Hartmann
    number~\cite{TEO11}.}
\label{table:m3}
\scalebox{1.}{
\begin{tabular}{llll}    
\hline\noalign{\smallskip}
type of flow & Branch 1~~~        & Branch 2~~~                    & Branch 3 \\
\noalign{\smallskip}\hline\noalign{\smallskip}
P.O.         & $(3.95,12.2)~~~~~$ &                                &   \\
2T           & $(3.39,3.95)~~~~~$ & $(0.85,2.45)$                  & $[0,2.9)$ \\
3T           & $(3.05,3.39)~~~~~$ & $[0,0.85)\cup(2.45,3.37)~~~~~$ &   \\
\noalign{\smallskip}\hline
\end{tabular}}
\end{table}

\subsection{Flows with $m=1,~m_{\textnormal{max}}=2$}
\label{sec:m2}

\begin{figure*}[h!]
\begin{center}
  \includegraphics[scale=1.35]{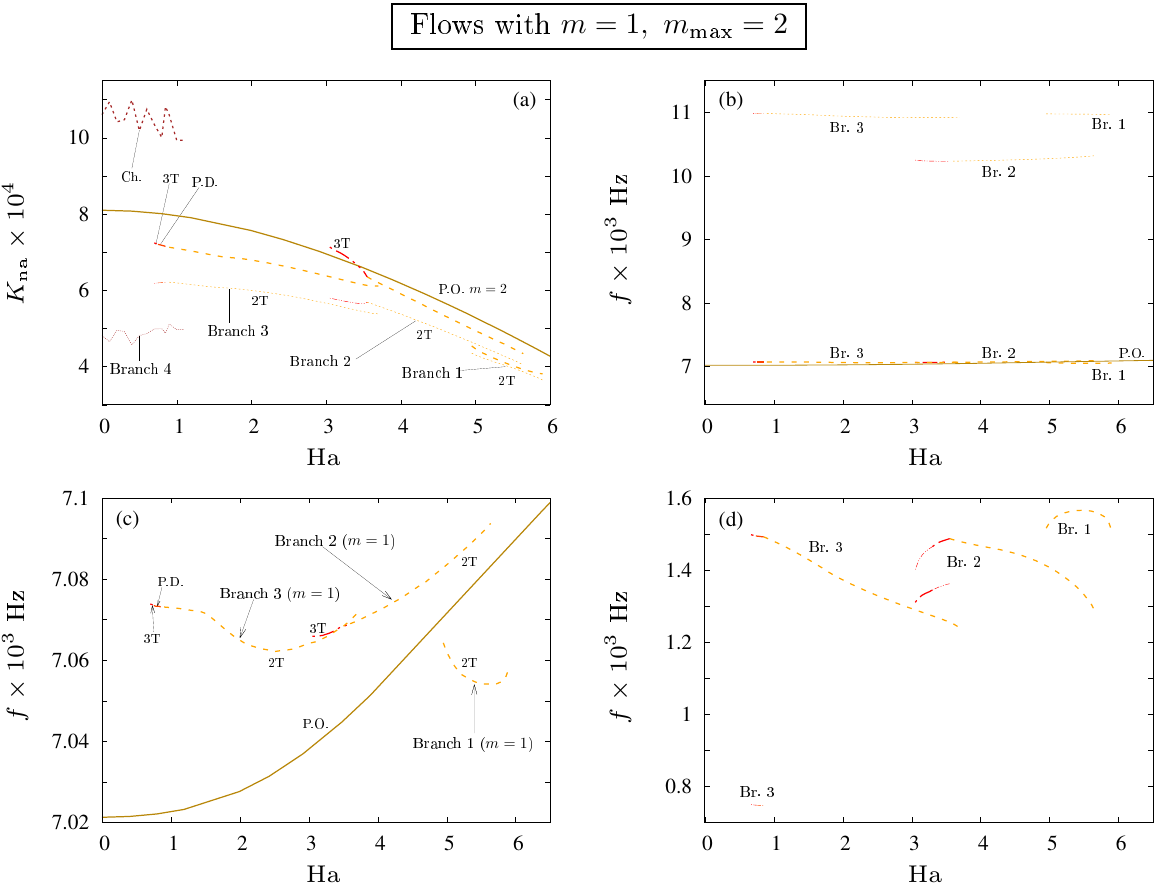}
\end{center}
\caption{Flows with $m=1$ and $m_{\text{max}}=2$. (a) Maximum (thick
  line) and minimum (thin line) of $K_{\text{na}}$. (b) 1st (thick
  line) and 2nd (thin line) dominant frequencies (the two having
  largest amplitude) of the flow. (c) Detail of (b) showing the
  frequencies in the range of $\approx 0.007$ Hz. (d) 1st (thick line)
  and 2nd (thin line) dominant frequencies of the volume-averaged
  kinetic energy.  The type of solutions --periodic orbit (P.O., solid
  line), invariant tori (2T, dashed line), invariant tori with period
  doubling (P.D., dashed-dotted line), invariant tori with three
  frequencies (3T, dashed-double-dotted line), and chaotic flows (Ch.,
  dotted line)-- is marked on each branch.}
\label{fig:bif_diagr_m2}    
\end{figure*}

This section focuses on the analysis of the class of flows with
$m=1,~m_{\text{max}}=2$, previously shown in
Fig.~\ref{fig:bif_diagr}. Figure~\ref{fig:bif_diagr_m2}, displaying
the same quantities as Fig.~\ref{fig:bif_diagr_m3}, summarises this
analysis. Several isolated branches (Branch 1, 2, 3, and 4) are
displayed and the flow on each branch can exhibit different types of
temporal dependence. On Branch 1 only invariant 2T flows are obtained
(right bottom corner, Fig.~\ref{fig:bif_diagr_m2}) while Branch 4 is
formed only by strongly oscillatory chaotic flows. Branches 2 and 3
contain 2T solutions as well as a small interval of 3T solutions.

The curves of 2T solutions in the Branches 1 and 2 were studied
previously in Ref.~\citen{GSGS19} and the spatio-temporal symmetries
of the solutions were determined. These branches were obtained by
taking initial conditions, and perturbing them, on the branch of
rotating waves (P.O.) with $m=2$-fold azimuthal symmetry first
computed in Ref.~\citen{GaSt18}. By slightly decreasing $\Ha$ and
taking initial conditions on the left part of the curve of 2T
solutions on Branch 2 (at around $\Ha\approx 3.6$) a branch of 3T
solutions is found. By further decrease of $\Ha$, and similarly to
what occurred on Branch 1 for flows with $m_{\text{max}}=3$ (previous
section), the 3T solutions on Branch 2 give rise to 2T solution on the
new Branch 3 and a hysteretic behaviour is found. This happens at around
$\Ha\approx 3$. As commented in the previous section, this behaviour
may be associated to a subcritical pitchfork bifurcation on Branch 2.

By decreasing $\Ha$ the curve of 2T solutions on Branch 3 extends down
to $\Ha\approx 0.9$ and undergoes a period doubling bifurcation
followed by a Hopf bifurcation giving rise to 3T solutions. This
occurs in a very narrow interval $\Ha\in(0.7,0.9)$. For $\Ha<0.7$ the
only flows found are the chaotic ones of Branch 4. Again, hysteresis
occurs and the presence of chaotic flows extends up to $\Ha\approx
1$. The curves of the maximum and minimum volume-averaged
nonaxisymmetric ($m\ne0$) kinetic energy of flows belonging to Branch
4 (shown in Fig.~\ref{fig:bif_diagr_m2}) are clearly non-smooth, which
is a sign of the chaotic behaviour of the solutions. Further
confirmation of this chaotic time dependence will be provided in the
next section when comparing Branch 4 with the chaotic flows found when
the azimuthal symmetry is $m=2$.

Figures~\ref{fig:bif_diagr_m2}(b,c,d) correspond to the frequency
analysis of regular solutions, i.\,e. of Branches 1, 2, and
3. Similarly as described in the previous section, for flows with
$m_{\text{max}}=3$, the main frequency is very close to the drifting
frequency (around $7$ mHz) of the rotating waves (P.O.) (see
Fig.~\ref{fig:bif_diagr_m2}(b) and its detail (c)). The second
dominant frequency seen in Fig.~\ref{fig:bif_diagr_m2}(b) is related
to the frequency giving rise to the quasiperiodic behaviour,
physically corresponding to the modulation of the wave, i.\,e., to the
oscillation of volume averaged properties. As displayed in
Fig.~\ref{fig:bif_diagr_m2}(d), the latter oscillations have large
temporal time scales when compared with the time scales associated
with the azimuthal drifting of the wave. Notice that for the invariant
tori with period doubling P.O. of Branch 3 the main frequency and the
halved frequency are both shown on the left part of
Fig.~\ref{fig:bif_diagr_m2}(d).

\begin{figure*}[h!]
\begin{center}
  \includegraphics[scale=1.15]{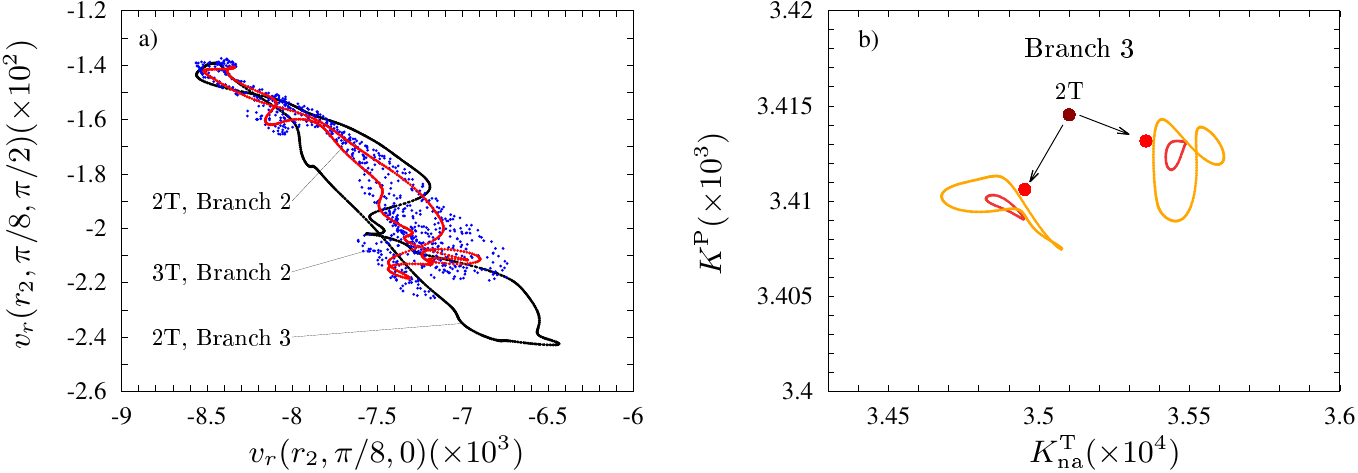}
\end{center}
\caption{Flows with $m=1,~m_{\text{max}}=2$ (a) Poincar\'e sections
  defined by $v_r(r_1,\pi/8,0)=-6\times 10^{-2}$. The radial positions
  are $r_1=r_{\mathrm{i}}+0.5d$ and
  $r_2=r_{\mathrm{i}}+0.85d$. Invariant tori with two frequencies (2T)
  of the 2nd and 3rd branches at $\Ha=3.7$ as well as an invariant
  torus with 3 frequencies (3T) of the 2nd branch at $\Ha=3.5$ are
  displayed. (b) Poincar\'e sections defined by the constraint
  $K(t)=\overline{K}$, $K$ being the volume-averaged kinetic energy
  and $\overline{K}$ its time average. The volume-averaged poloidal
  kinetic energy $K^{\text{P}}$ is displayed versus the
  volume-averaged toroidal nonaxisymmetric energy
  $K^{\text{T}}_{\text{na}}$. On the 3rd branch, invariant tori with
  two frequencies at $\Ha=1$ and at $\Ha=0.8$ are marked with large
  points. Notice that from $\Ha=1$ to $\Ha=0.8$ a period doubling
  bifurcation (see arrows) has occurred. For smaller $\Ha$ a
  transition to 3 frequency invariant tori (small red at $\Ha=0.74$
  and large orange at $\Ha=0.7$ closed curves) occurs.}
\label{fig:poinc_m2}    
\end{figure*}

Several Poincar\'e sections are displayed on Fig.~\ref{fig:poinc_m2}
to demonstrate the type of time dependence of flows with $m=1$ and
$m_{\text{max}}=2$. On Fig.~\ref{fig:poinc_m2}(a) the Poincar\'e
sections are defined by $v_r(r_1,\pi/8,0)=-6\times 10^{-2}$ whereas on
Fig.~\ref{fig:poinc_m2}(b) the sections are defined by the constraint
$K(t)=\overline{K}$, $K$ being the volume-averaged kinetic energy and
$\overline{K}$ its time average. As commented previously, considering
a volume averaged quantity (such as $\overline{K}$) in the definition
of the Poincar\'e section allows to identify 3 frequency solutions
(3T), as they correspond to closed curves (the drifting frequency is
removed by volume averaging).


Figure~\ref{fig:poinc_m2}(a) displays the sections of three different
flows: 2T solution ($\Ha=3.7$) and 3T solution ($\Ha=3.5$) on Branch 2
and a 2T solution ($\Ha=3.7$) on Branch 3. As the radial velocity
$v_r$ is not a volume-averaged quantity, the 2T solutions correspond
to closed curves and the 3T solutions correspond to the cloud of
points. The shape of the curves is indeed characteristic of complex 2T
attractors. Comparing these two curves at the same $\Ha=3.7$ helps to
compare Branch 2 with Branch 3. Notice that oscillations of $v_r$ are
more developed on Branch 3 as the Poincar\'e section encloses a larger
area. Between $\Ha=1$ and $\Ha=0.8$ a period doubling of 2T flows on
Branch 3 can be identified from Fig.~\ref{fig:poinc_m2}(b). At $\Ha=1$
the Poincar\'e section is a single point whereas at $\Ha=0.8$ it
consists of two points, i.\,e. it has doubled the period. In addition,
at $\Ha=0.74$ and $\Ha=0.7$ a new independent frequency appears -by
means of a Hopf bifurcation- giving rise to 3T solutions whose
Poincar\'e sections of volume-averaged quantities become closed
curves.

\begin{table}[t]
\caption{Range of Hartmann numbers for the stability of the different
  types of temporal dependence along the different branches for flows
  with $m=1$ and $m_{\text{max}}=2$. The interval of definition of the
  nonaxisymmetric radial jet instablitity is $\Ha\in[0,\Ha_c)$ with
    $\Ha_c=12.2$ being the critical Hartmann number~\cite{TEO11}.}
\label{table:m2}
\scalebox{1.}{
\begin{tabular}{lllll}    
\hline\noalign{\smallskip}
type of flow~~~~~ & Branch 1~~~        & Branch 2~~~        & Branch 3~~~       & Branch 4 \\
\noalign{\smallskip}\hline\noalign{\smallskip}
2T              & $[4.95,5.9]~~~~~$  & $[3.55,5.64]~~~~~$ & $[0.75,3.7]~~~~~$ & \\
3T           &                    & $[3.05,3.55]~~~~~$ & $[0.7,0.74]~~~~~$ & \\
Ch.          &                    &                    &                   & $~~~[0,1]$\\
\noalign{\smallskip}\hline
\end{tabular}}
\end{table}

With the exception of the branch of P.O., all flows described in this
section are stable. The interval of Hartmann numbers for the stability
of each type of temporal dependence (2T, 3T, and Ch.) on each branch
shown in Fig.~\ref{fig:bif_diagr_m2}(a) is provided in
Table~\ref{table:m2}.

\subsection{Flows with $m=2$}
\label{sec:md2}

\begin{figure*}[h!]
\begin{center}
  \includegraphics[scale=1.15]{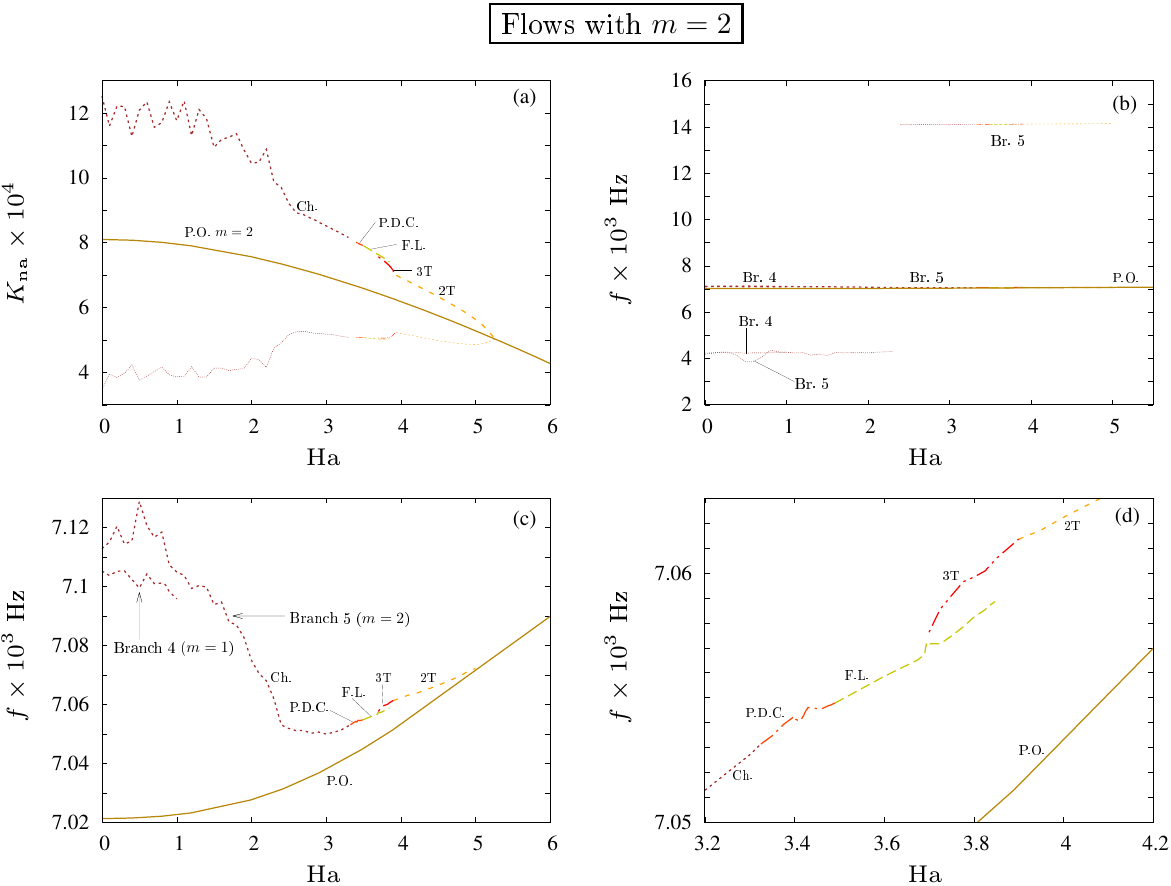}
\end{center}
\caption{Flows with $m=2$. (a) Maximum (thick line) and minimum (thin
  line) of $K_{\text{na}}$. (b) 1st (thick line) and 2nd (thin line)
  dominant frequencies (the two having largest amplitude) of the
  flow. (c) Detail of (b) showing the frequencies in the range of
  $\approx 0.007$ Hz. (d) 1st (thick line) and 2nd (thin line)
  dominant frequencies of the volume-averaged kinetic energy.  The
  type of solutions --periodic orbit (P.O., solid line), invariant
  tori (2T, dashed line), invariant tori with period doubling (P.D.,
  dashed-dotted line), invariant tori with three frequencies (3T,
  dashed-double-dotted line), and chaotic flows (Ch., dotted line)--
  is marked on each branch.}
\label{fig:bif_diagr_md2}    
\end{figure*}

In this section the class of flows with $m=2$, first displayed on
Fig.~\ref{fig:bif_diagr} and obtained imposing $m=2$-fold azimuthal
symmetry on the DNS, is
investigated. Figure~\ref{fig:bif_diagr_md2}(a) displays the maximum
and minimum values of the volume-averaged nonaxisymmetric $m\ne 0$
kinetic energy of the different types of flows. According to
Refs.~\citen{GaSt18,GSGS19} a branch of 2T solutions is born at
$\Ha\approx 5.25$ through a Hopf bifurcation on the unstable branch of
periodic orbits P.O.. This branch of 2T solutions extends down to
$\Ha\approx 3.94$ where a tertiary Hopf bifurcation gives rise to a
new branch of 3T solutions. A rich variety of flow types is found by
decreasing $\Ha\in[3,4]$ from this point. First, an interval of
frequency locking (F.L.) --a parameter region where the ratio of two
fundamental frequencies becomes a constant rational number (see
Ref.~\citen{SNGS04b} for instance)-- on a branch of 3T solutions is
found. This is followed by a doubling period cascade giving rise to
chaotic flows at the smallest $\Ha$.

The frequency analysis for the radial velocity summarised in
Fig.~\ref{fig:bif_diagr_md2}(b,c,d) provides further evidence for the
above mentioned scenario. Panel (b) indeed confirms what was observed
in Secs.~\ref{sec:m3}-\ref{sec:m2}: The main frequency of the flow is
quite constant and remains very close to the one corresponding to the
unstable rotating waves with $m=2$ at the same range of
parameters. The label Br.~5 on Fig.~\ref{fig:bif_diagr_md2}(b) (and on
its detail, Fig.~\ref{fig:bif_diagr_md2}(c)) refers to Branch 5, which
is the branch corresponding to $m=2$ displayed on
Fig.~\ref{fig:bif_diagr_md2}(a). The label Br.~4 corresponds to Branch
4, a branch of chaotic flows with $m=1,~m_{\text{max}}=2$ already
described in Sec.~\ref{sec:m2}. As happens for the nonaxisymmetric
kinetic energy, see Fig.~\ref{fig:bif_diagr} and compare
Fig.~\ref{fig:bif_diagr_m2}(a) with Fig.~\ref{fig:bif_diagr_md2}(a),
the values of the 1st and 2nd frequencies are very close giving
evidence of the relation between Branch 4 and Branch 5. Solutions
lying on the latter branch seem to be the projection onto the $m=2$
subspace of the corresponding solution with $m=1$, as this is slightly
more energetic (see Fig.~\ref{fig:bif_diagr}) because it contains more
spherical harmonics.

Figure~\ref{fig:bif_diagr_md2}(d) details the bifurcation scenario in
the interval $\Ha\in[3.2,4.2]$. As commented before, an interval of
frequency locking on a branch of 3T solutions is found for
$\Ha\in[3.5,3.8]$. In this range the ratio is $f_1/f_2=1.5\pm 3\times
10^{-3}$, $f_1,f_2$ being the 1st and 2nd dominant Laskar~\cite{Las93}
frequencies of the volume averaged kinetic energy spectrum. In a
narrow interval around $\Ha>3.4$ a sequence of period doubling
bifurcations is found. The estimated critical parameters, $\Ha_{i}$
($i=1,..,4$), are $\Ha_1=3.491$, $\Ha_2=3.423$, $\Ha_3=3.4073$, and
$\Ha_4=3.4039$. The Feigenbaum iterates are $\delta_1=4.33$ and
$\delta_2=4.62$ (with the definition
$\delta_i=(\Ha_{i+1}-\Ha_{i})/(\Ha_{i+2}-\Ha_{i+1})$), in good
agreement with the Feigenbaum constant $\delta=4.6692$. A period
doubling cascade was also found in Ref.~\citen{GSDN15} when
restricting the azimuthal symmetry in the context of thermal
convection in rotating spherical shells.

\begin{figure*}[h!]
\begin{center}
  \includegraphics[scale=1.15]{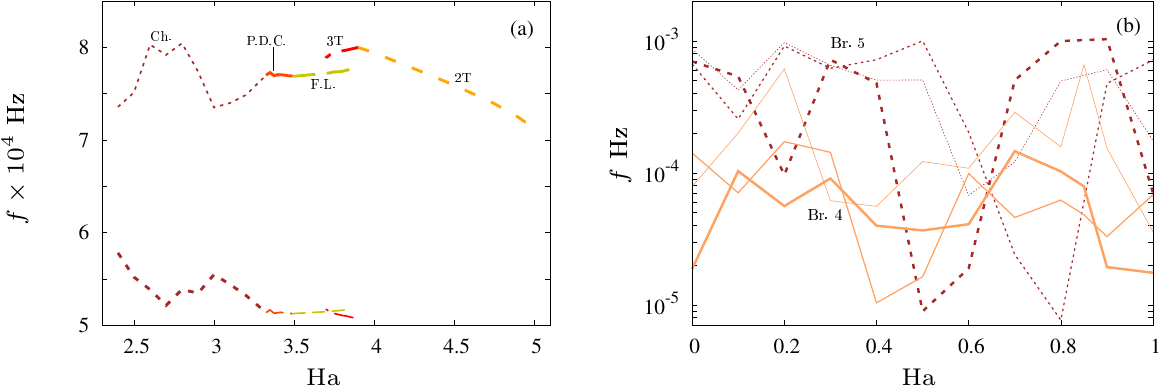}
\end{center}
\caption{Two different branches with $m=1$-fold azimuthal symmetry
  (Branch 4) and with $m=2$-fold azimuthal symmetry (Branch 5) are
  shown.  (a) the 1st (thick line) and 2nd (thin line) dominant (those
  of largest amplitude) frequencies of a time series of the
  volume-averaged kinetic energy for solutions of Branch 5. (b) As (a)
  but including the chaotic solutions with $m=1$-fold azimuthal
  symmetry (Branch 4) in the range $\Ha\in[0,1]$. The type of
  solutions --invariant tori with two frequencies (2T, dashed line),
  invariant tori with three frequencies (3T, dashed-double-dotted
  line), invariant tori with two frequencies coming from period
  doubling cascade (P.D.C., dashed-dotted), chaotic flows (Ch., dotted
  line) and periodic orbit (P.O., solid line) -- is marked on each
  branch. For this class of flows an interval of frequency locking on
  a branch of 3T solutions is obtained (F.L., long-dashed line).}
\label{fig:bif_diagr_md2_c6var}    
\end{figure*}

Figure~\ref{fig:bif_diagr_md2_c6var}(a,b) displays the main
frequencies associated with the time series of the volume-averaged
kinetic energy. It supports the visualisation of long time scales of
the flow evolution (with frequencies less than a mHz) associated with
the modulation. Figure~\ref{fig:bif_diagr_md2_c6var}(a) evidences the
region of frequency locking as the curves for the 1st and 2nd
frequencies (thick/thin line) have the same shape and we have checked
that they differ only by a constant ($3/2$)
factor. Figure~\ref{fig:bif_diagr_md2_c6var}(b) displays the three
main frequencies only for the chaotic flows of branches 4 and 5 in the
range $\Ha\in[0,1]$. The chaotic behaviour of solutions on each branch
gives rise to sharp changes as $\Ha$ is varied. All the frequencies
are very small giving rise to time scales up to $10^5$ seconds,
i.\,e. a whole day. We note that both, Branch 4 and 5, exhibit the
same behaviour with $\Ha$ providing further confirmation of the
relation among both branches.

\begin{figure*}[h!]
\begin{center}
  \includegraphics[scale=1.15]{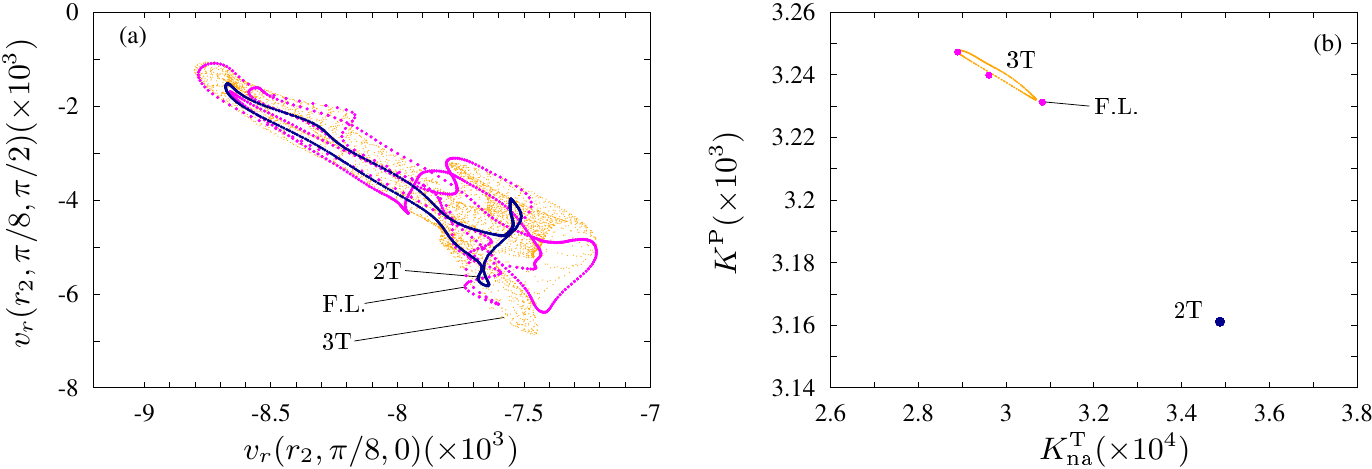}
\end{center}
\caption{Flows with $m=2$ (a) Poincar\'e sections defined by
  $v_r(r_1,\pi/8,0)=-6\times 10^{-2}$. The radial positions are
  $r_1=r_{\mathrm{i}}+0.5d$ and $r_2=r_{\mathrm{i}}+0.85d$. An
  invariant torus with two frequencies (2T) at $\Ha=4$ as well as an
  invariant torus with 3 frequencies (3T) and a resonant (frequency
  locked) 3T solution at $\Ha=3.8$ are displayed. (b) Poincar\'e
  sections defined by the constraint $K(t)=\overline{K}$, with $K$
  being the volume-averaged kinetic energy and $\overline{K}$ its time
  average. The volume-averaged poloidal kinetic energy $K^{\text{P}}$
  is displayed versus the volume-averaged toroidal nonaxisymmetric
  energy $K^{\text{T}}_{\text{na}}$. The invariant torus with two
  frequencies at $\Ha=4$ is marked with a large point. The closed
  curve corresponds to the 3T solution at $\Ha=3.8$. The set of three
  points lying on the curve correspond to the resonant 3T solution at
  $\Ha=3.8$. }
\label{fig:poinc_md2}    
\end{figure*}

Again, confirmation of the bifurcation scenario is provided by the
inspection of the Poincar\'e sections. Figure~\ref{fig:poinc_md2}(a)
provides a picture of the section, defined in terms of the radial
velocity, of a 3T solution (cloud of dots) and the corresponding
section of a frequency locking solution (closed larger curve) at
$\Ha=3.8$. Both sections are similar (same elongation and similar
occupation of area) with that of the 2T solution at $\Ha=4$, meaning
that the three solutions are related as they arise at different
bifurcations on the same branch. If the volume averaged kinetic energy
is used to define the Poincar\'e section (Fig.~\ref{fig:poinc_md2}(b))
the situation is even more clear since the 3T solution is a closed
curve and the frequency locked 3T solution is a set of 3 points.

\begin{figure*}[h!]
\begin{center}
  \includegraphics[scale=1.15]{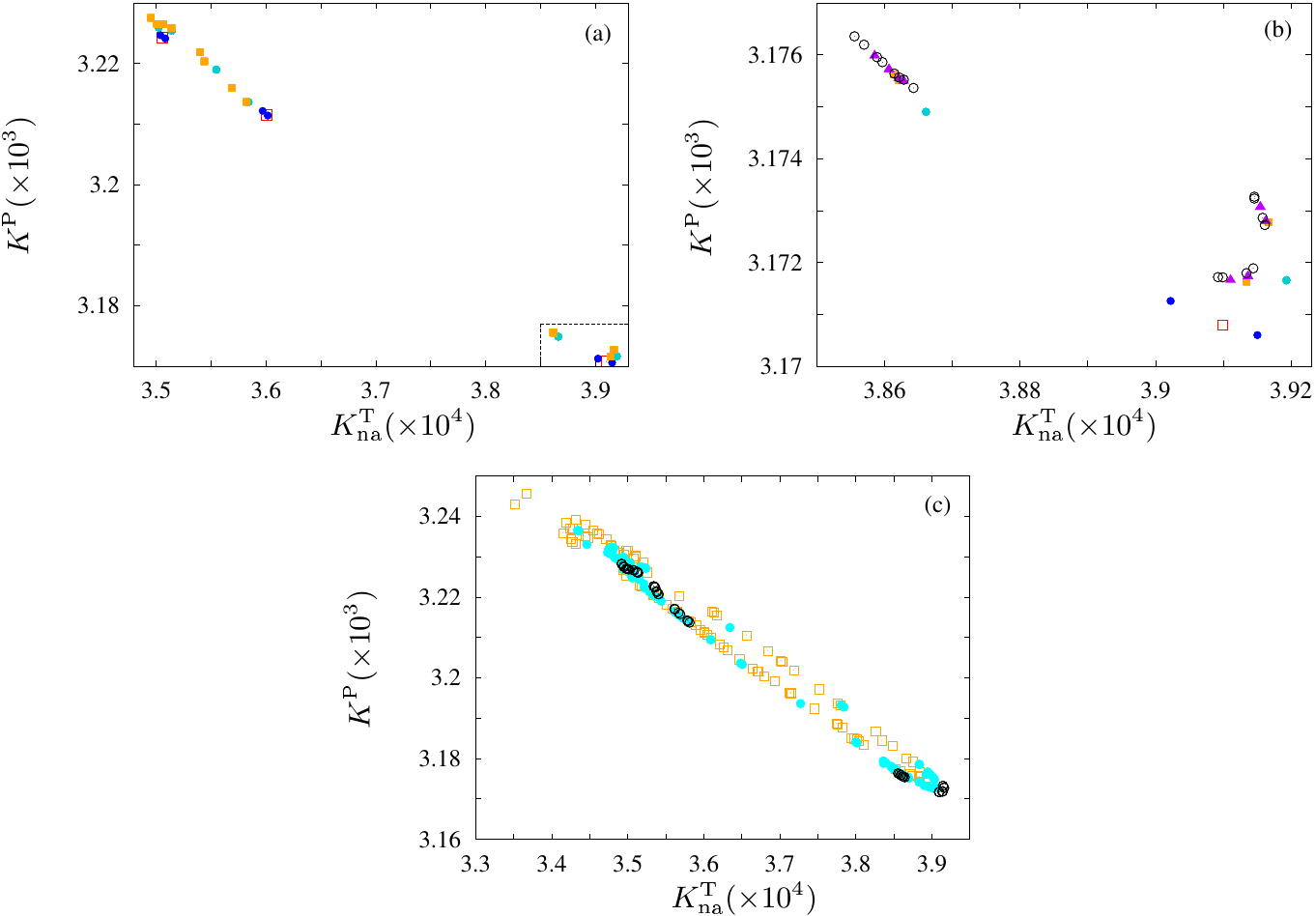}
\end{center}
\vspace{-8.mm}
\caption{Period doubling cascade with $m=2$. The Poincar\'e sections
  are as in Fig.~\ref{fig:poinc_m3}. (a) The set of three points (open
  squares) correspond to a resonant 3T solution at $\Ha=3.4925$. At
  $\Ha=3.49$ and $\Ha=3.43$ the Poincar\'e sections contain 6 points
  (full circles). At $\Ha=3.41$ the Poincar\'e section contains 12
  points (full squares). (b) Detail of (a) showing additional
  Poincar\'e sections at $\Ha=3.406$ (24 points, full triangles) and
  at $\Ha=3.403$ (48 points, open circles). (c) Period doubling flow
  at $\Ha=3.403$ (48 points, open circles) and chaotic flows at
  $\Ha=3.35$ (full circles) and $\Ha=3.3$ (open squares).}
\label{fig:poinc_md2_perdoub}    
\end{figure*}

Figure~\ref{fig:poinc_md2_perdoub} displays the Poincar\'e sections
defined in terms of volume averaged properties (see figure caption) of
flows undergoing a sequence of period doubling bifurcations. These
Poincar\'e sections are again quite useful to identify the period
doubling cascade, on a branch of 3T solutions with frequency locking,
as the number of points is doubled at each bifurcation. The first and
second period doubling bifurcations are visualised on
Fig.~\ref{fig:poinc_md2_perdoub}(a). The set of 3 points at
$\Ha=3.4925$ (open squares) become 6 points (full circles) at
$\Ha=3.49,3.43$ and 12 points (full squares) at
$\Ha=3.41$. Figure~\ref{fig:poinc_md2_perdoub}(b) deepens this
investigation providing the sections for solutions at $\Ha=3.406$ (24
full triangles) and $\Ha=3.403$ (48 open circles) beyond the 3rd and
4th bifurcations. At $\Ha=3.35$ and $\Ha=3.3$ chaotic flows are
compared with the last regular solution of the Feigenbaum cascade in
Fig.~\ref{fig:poinc_md2_perdoub}(c).

\begin{figure*}[h!]
\begin{center}
  \includegraphics[scale=1.15]{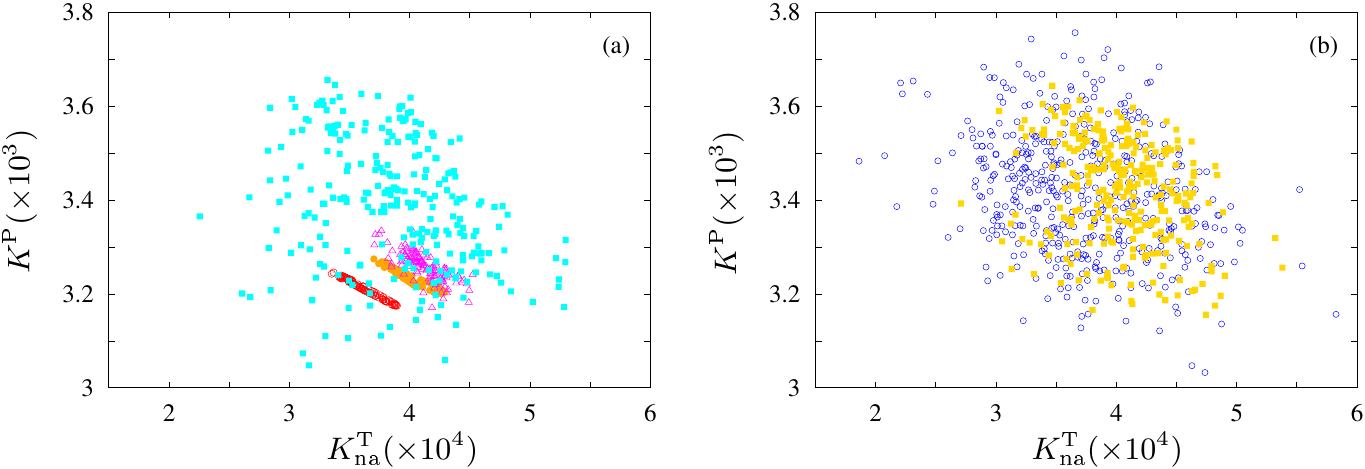}
\end{center}
\vspace{-8.mm}
\caption{Chaotic flows with $m=2$ and with
  $m_{\text{max}}=2,~m=1$. The Poincar\'e sections are as in
  Fig.~\ref{fig:poinc_m3}. (a) Chaotic flows with $m=2$ at $\Ha=3.3$
  (open circles), $\Ha=3$ (full circles), $\Ha=2.5$ (open triangles)
  and at $\Ha=1$ (full squares). (b) Comparison of chaotic purely
  hydrodynamic flows ($\Ha=0$) with $m=2$ (open circles) and with
  $m_{\text{max}}=2,~m=1$ (full squares).}
\label{fig:poinc_md2_m1_chaos}    
\end{figure*}

Finally, Fig.~\ref{fig:poinc_md2_m1_chaos} investigates the $\Ha$
dependence of the Poincar\'e sections for chaotic flows down to
$\Ha=0$. Volume averaged properties are considered as in
Fig.~\ref{fig:poinc_md2_perdoub}. Figure~\ref{fig:poinc_md2_m1_chaos}(a)
compares chaotic flows with $m=2$ arising from the period doubling
cascade to those with $m=2$ more developed at lower $\Ha$. By
decreasing the magnetic field strength the chaotic flows become more
oscillatory (and thus the cloud of points spreads over larger
areas). The agreement between the Poincar\'e sections of flows
belonging to Branch 5 (with $m=2$) and those belonging to Branch 4
(with $m=1,~m_{\text{max}}=2$) is quite noticeable as seen on
Fig.~\ref{fig:poinc_md2_m1_chaos}(b) displaying the Poincar\'e
sections at $\Ha=0$. In agreement with Fig.~\ref{fig:bif_diagr_md2}(a)
the chaotic flow with $m=2$ becomes slightly more oscillatory than the
corresponding flow with $m=1$. For chaotic solutions on Branch 5 the
number of spherical harmonics amplitudes is halved (as the azimuthal
symmetry is $m=2$) with respect to those on Branch 4 (with azimuthal
symmetry $m=1$). As flows on Branch 4 and Branch 5 have comparable
energy (shown on Fig.~\ref{fig:bif_diagr}), this energy is shared out
by more modes $m$ in the case of Branch 4, which may facilitate the
damping of the oscillations. The kinetic energy contribution of the
$m=2$ mode is comparable for both branches and the contribution of
each of the remaining modes ($m=1,3,4,5,6,8$ for Branch 4 and
$m=4,6,8$ for Branch 5) is similar and roughly one order of magnitude
smaller.

The flows with $m=2$ considered in this section are all unstable. This
has been checked by adding an $m=1$ random perturbation which always
grows, even with very small amplitude ($O(10^{-12})$). Perturbing the
chaotic flows for $\Ha\in[0,1]$ with $m=2$ gives rise to the chaotic
flows with $m=1$ lying on Branch 4. Because chaotic flows with $m=2$
and with $m=1$ have very similar kinetic energies and frequencies (see
Fig.~\ref{fig:bif_diagr_md2_c6var}(b)) they are clearly related and
thus Branch 4 seems to be related with Branch 5, that is, the unstable
branch with $m=2$.

\subsection{Flow patterns}
\label{sec:pat}

\begin{figure*}[h!]
\begin{center}
  \includegraphics[scale=2.2]{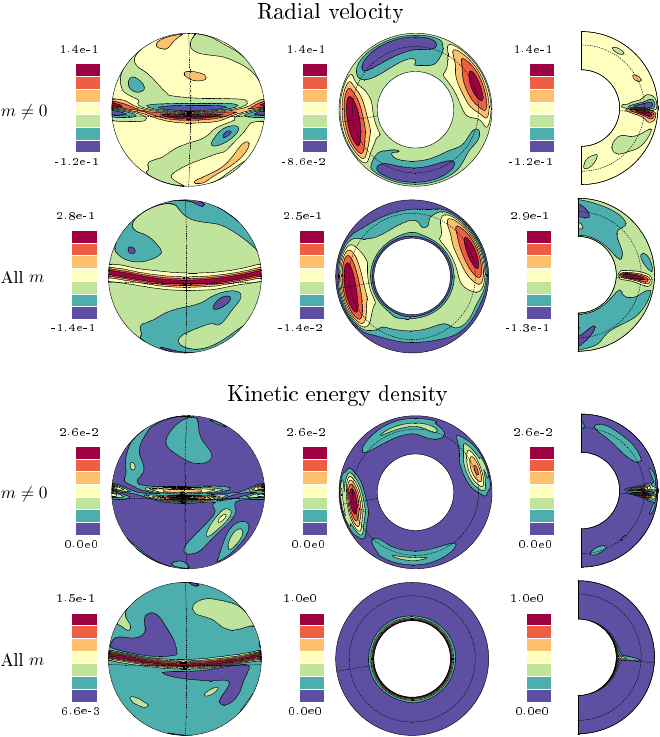}
\end{center}
\caption{Contour plots for a chaotic flow with $m=1,~m_{\text{max}}=2$
  at $\Ha=0$. 1st/2nd rows: nonaxisymmetric and total components of
  the radial velocity. 3rd/4th rows: nonaxisymmetric and total
  components of the kinetic energy. Spherical, colatitudinal, and
  meridional sections are displayed on each row. For the
  nonaxisymmetric component every section is located in a such way
  that it is cutting a relative maxima of the respective field. The
  sections for the total component are located at the same positions
  than those of the nonaxisymmetric component to facilitate the
  comparison between both components. The position of the sections is
  marked with a dashed line.}
\label{fig:pat}    
\end{figure*}

In this section the flow patterns for a chaotic solution are briefly
described. The flow patterns do not change substantially for
$\Ha\in[0,12.2]$. A comprehensive description of the topology of
quasiperiodic waves, around $\Ha\sim 5$, was performed in
Ref.~\citen{GSGS19} so only few details are provided in this
section. The studied range corresponds to the radial jet
instability~\cite{HJE06}, which is nonaxisymmetric and asymmetric with
respect the equatorial plane. A chaotic solution at $\Ha=0$ with $m=1$
and $m_{\text{max}}=2$ is considered. The contour plots of the
nonaxisymmetric and total components of the radial velocity are
displayed in the first and second rows of Fig~\ref{fig:pat}. The third
and fourth rows are analogous but for the kinetic energy. Three
different sections -spherical, colatitudinal, and meridional- are
selected on each row passing through a maximum of the corresponding
field so the global picture emerges.

Quite regular nonaxisymmetric patterns are observed (1st/3rd rows)
although the solution is chaotic. They compare very well with the
patterns of quasiperiodic solutions studied in Ref.~\citen{GSGS19}
(see for instance their Fig. 6). The flow consist of two main cells,
as $m_{\text{max}}=2$, but it is not invariant under $\pi$ azimuthal
rotations as it retains the $m=1$ azimuthal symmetry (best shown on
3rd row). As it is usual for the spherical Couette system, with
moderate $\Ree=10^3$, the main flow is nearly axisymmetric consisting
of a strong jet that is located close to equatorial latitudes (see
meridional sections on 2nd/4th rows of Fig~\ref{fig:pat}) and extends
radially outwards from the inner sphere.

\clearpage
\section{Discussion and Conclusions}
\label{sec:disc}

The present study constitutes a sequel of two previous papers
(Refs.~\citen{GaSt18} and~\citen{GSGS19}) where the dependence on the
Hartmann number of rotating waves (periodic orbits) and modulated
waves (invariant tori), respectively, were investigated in detail for
the azimuthal symmetries $m=2,3,4$ and for fixed $\Ree=10^3$ and
aspect ratio $\chi=0.5$. In this new work we infer the existence of
flows with complex temporal dependence arising from the branches of
regular solutions previously computed in Refs.~\citen{GaSt18,GSGS19}
in the range $\Ha\in[0,6]$.

By means of direct numerical simulations we have been able to obtain
several branches of modulated waves. Commonly, these solutions are
invariant tori having two fundamental frequencies as in
Ref.~\citen{GSGS19} or~\citen{GNS16} (in the context of rotating
thermal convection) but we have found modulated waves having 3
independent frequencies. According to Ref.~\citen{NRT78}, any small
smooth perturbation applied to these 3-frequency solutions should
provide a strange chaotic attractor, but it is not clear if this will
happen in non-ideal physical systems~\cite{GOY83}. We have found, in
agreement with Ref.~\citen{GOY83}, that three frequency solutions (3T)
are typical in the magnetised spherical Couette system.

We have classified the flows by analysing the azimuthal symmetry $m$
and the most energetic azimuthal wave number $m_{\text{max}}$. Three
different classes of flows, with $m_{\text{max}}=3$, with
$m=1,~m_{\text{max}}=2$ and with $m=2$, are found and the
bifurcations occurring on each of these classes are analysed. A rich
variety of bifurcations and types of solutions have been found,
including hysteretic behaviour. Secondary as well as tertiary Hopf
bifurcations, period doublings, period doubling cascade and frequency
locking are the typical mechanisms found to change the
dynamics. Thanks to this rich dynamics several regions of
multistability, comprising two, three and even four different types of
solutions exist.

The appearance of multiple states in experimental flows strongly
depends on the conditions assumed for the initial state~\cite{NaTs95},
and in DNS the type of perturbation applied determines the type of
mode that will be selected among the bifurcated solutions. Ultimately,
the geometry of the stable invariant manifolds and their folding in
the phase space determines the basin of attraction of each regular
solution. In this way, the computation of these manifolds could help
to understand why a given initial condition leads to the $m=1,
m_{\text{max}}=2$ or to the $m=1, m_{\text{max}}=3$ branch,
respectively.

Table~\ref{table:percent} provides a quantitative overview of the
different types of solutions found. For $\Ree=10^3$ and $\chi=0.5$ the
nonaxisymmetric radial jet instability extends through
$\Ha\in[0,12.2]$ where $12.2$ is the critical Hartmann number. This
was found thanks to the linear stability analysis of the basic flow of
Ref.~\citen{TEO11}. Then, the ratio of the interval of stability of a
given type of flow to the total interval ($[0,12.2]$ of the definition
of the radial jet instability) provides a measure of the probability
to obtain the given flow from a randomly distributed initial
$\Ha\in[0,12.2]$. As it is shown in Table~\ref{table:percent} a rather
large probability of $1/5$ is expected for the case of three frequency
solutions. This means that 3 frequency solutions are quite common in
the spherical Couette system.

The comprehensive analysis presented here is of fundamental importance
for current and future comparisons with the HEDGEHOG experiment, which
is designed to run in a quasi-laminar regime ($\Ree\in[10^3,10^4]$ and
$\Ha<10^3$). The present study has considered a single $\Ree=10^3$ as
our focus was to study the dependence on the magnetic field
strength. The analysis of the sensitivity to $\Ree$ requires further
research, especially for $\Ree\sim 10^4$. Previous numerical
simulations (e.g., Ref.~\citen{GJG11}) have shown that the effect of
increasing $\Ree$ is to favour the appearance of flows with
$m_{\text{max}}=2$ and, eventually, to develop turbulent
motions. According to our results, modulated waves at $\Ha<5$ could be
obtained in the experimental setup as they are the only stable and
thus physically realisable flows. The frequency analysis we have
performed has demonstrated that the frequency associated with the
azimuthal drift of the flow is quite robust and remains nearly
constant for $\Ha\in[0,6]$ and for each class of flows. For each
complex wave this frequency is quite similar to that of the unstable
rotating wave with $m=m_{\text{max}}$ at the same $\Ha$.

Moreover, the frequency analysis provides the time scales of the
modulation thus constraining the required duration of the related
experimental runs. For instance, in the case of flows with
$m=1,~m_{\text{max}}=2~~~$ the main frequency of modulation for 2T
solutions at $\Ha<6$ is smaller than 2 mHz, which requires time scales
larger than 10 min. Then, the experiment should run during almost a
day to capture at least 100 modulation periods. Very long experimental
runs should be avoided because of the signal quality of the ultrasonic
Doppler velocimetry (UDV) measurements degrades over time (see
Ref.~\citen{KKSS17} for details). The latter is caused by the
evolution of tracer particles (oxydes, solid binary phase combinations
of GaInSn) distribution within the liquid metal. The tracer particles
sediment during the experimental run so that the signal quality
decreases. We notice that relative variations of $\Ree$ and $\Ha$ due
to variations of the viscosity ($0.5\%$), conductivity ($0.1\%$) or
density ($0.01\%$) related with temperature changes (around $5$
degrees in the laboratory during a day) lead to a total variation of
the dimensionless parameters of about $\mathcal{O}(1\%)$. This is an
acceptable inconvenience as we assume a sufficiently smooth transition
for the bifurcation diagrams in the parameter space (see for instance
Fig.~\ref{fig:bif_diagr}), except close to bifurcation points.

\begin{table}[t]
\caption{Occurrence of the different types of temporal dependence. The
  interval of definition of the nonaxisymmetric radial jet
  instablitity is $\Ha\in[0,\Ha_c)$ with $\Ha_c=12.2$ being the
    critical Hartmann number~\cite{TEO11}. The percents represent the
    ratio between the length of the interval of stability of a given
    type of flow with respect the whole interval of definition of the
    radial jet instability.}
\label{table:percent}
\scalebox{1.}{
\begin{tabular}{lllll}    
\hline\noalign{\smallskip}
Flow class            & rotating wave~~~ & modulated wave 2T~~~ & modulated wave 3T~~~ & Chaos  \\
\noalign{\smallskip}\hline\noalign{\smallskip}
$m_{\text{max}}=3$      &  $~~~~67.6\%$     & $~~~~~~~~~~41.5\%$  & $~~~~~~~~~~17.3\%$   &         \\
$m=1,~m_{\text{max}}=2~~~$ &                & $~~~~~~~~~~49.1\%$  & $~~~~~~~~~~4.4\%$    & $~~8.2\%$ \\
\noalign{\smallskip}\hline
\end{tabular}}
\end{table}


%
%

%

\begin{acknowledgments}
F. Garcia kindly acknowledges the Alexander von Humboldt Foundation
for its financial support.  This project has also received funding
from the European Research Council (ERC) under the European Union’s
Horizon 2020 research and innovation programme (grant agreement No
787544).

\end{acknowledgments}


%

\end{document}